\documentclass [11pt]{article}
\usepackage{amsmath}
\usepackage{amssymb}
\usepackage{amscd}
\usepackage{amsthm}
\usepackage{amsopn}
\usepackage{xspace}
\usepackage{verbatim}
\usepackage{amsmath}
\usepackage{amsfonts}
\newtheorem{theorem}{Theorem}
\newtheorem{lemma}{Lemma}[section]
\newtheorem{proposition}{Proposition}[section]
\newtheorem{corollary}{Corollary}[section]
\newtheorem{definition}{Definition}[section]
\newtheorem{acknowledgment*}{Acknowledgment}
\newtheorem{remark}{Remark}[section]
\numberwithin{equation}{section}

\newcommand{\eps}{\varepsilon}
\newcommand{\R}{\mathbb R}

\newcommand{\oM}{\overline{ M}}

\newcommand{\oPsi}{\overline{\Psi}}
\newcommand{\tPsi}{\widetilde{\Psi}}
\newcommand{\tPhi}{\widetilde{\Phi}}
\newcommand{\ntPsi}{\widetilde{\Psi}^{(N)}}
\newcommand{\ntPhi}{\widetilde{\Phi}^{(N)}}
\newcommand{\orho}{\overline{\rho}}
\newcommand{\oV}{\overline{V}}
\newcommand{\ou}{\overline{u}}

\begin{document}

\title{Dynamics of a system of sticking
particles of a finite size on the line}
\author{\sc  Gershon Wolansky\\
{\small Department of Mathematics, Technion 32000, Haifa, ISRAEL}
\\
e.mail:gershonw@math.technion.ac.il }

\date{September 1, 2006}
\maketitle \vskip .3in \noindent AMS Subject classification:
35L65, 35L67


\begin{abstract}
The continuous limit of large systems of  particles of finite size
on the line is described. The particles are assumed to move freely
and stick under collision, to form compound particles whose mass
and size is the sum of the masses and sizes of the  particles
before collision, and whose velocity is determined by conservation
of linear momentum.
\end{abstract}

\section{Introduction}
\subsection{Review}
Models of  point particles on the line which stick under collision
have recently been considered in the literature, starting from the
pioneering paper of Zeldovich from  1970 (\cite{Z}, see also
\cite{SZ}). This model is extremely simple: Imagine a swarm of
point particles moving without interaction (constant velocity) on
the line. When two (or more) particles collide, they stick
together and continue to move at a constant velocity, determined
by conservation of their initial momentum.
\par Assuming for simplicity there are initially $N$ identical
particles of mass $1/N$, we can describe
 the density and momentum of this swarm by the fields
\begin{equation}\label{pointp} \rho_N=N^{-1}\sum_1^N \delta_{(x-x_i(t)) }, \ \
\ \ \rho_N u_N=  N^{-1}\sum_1^N v_i(t)\delta_{(x-x_i(t)) } \ ,
\end{equation} where here $\delta$ stands for the Dirak
delta-function. Here $x_i(t)$ is the position of the $i$-th
particle at time $t$ and $v_i(t)$ is its velocity at that
instance, assumed to be constant between collisions.  The
assumption of sticking collisions can be stated as
\begin{equation}\label{colpoint} v_i(t+)= \frac{\sum_{j; \ x_j(t)=x_i(t)}
v_j(t)}{\# \{ j; \ x_j(t)=x_i(t)\}} \  \ , x_i(t)=x_i(0)+\int_0^t
v_i (s)ds\ \ .  \end{equation}
\par
Let ($\rho_N,  u_N$), $N \rightarrow \infty$  be a sequence of
 the form  \eqref{pointp}. It is
said to converge weakly to
 ($\rho,u$) if
 \begin{equation}\label{weaklim} \lim_{N\rightarrow \infty} \int_{-\infty}^\infty \rho_N\phi dx=
 \int_{-\infty}^\infty
 \rho\phi dx\  , \ \lim_{N\rightarrow \infty} \int_{-\infty}^\infty \rho_N u_N\phi dx= \int_{-\infty}^\infty
 \rho u \phi  dx \ \end{equation}
 for any $\phi\in C_0(R)$ and $t\geq 0$.
 The first rigorous treatment of this limit   was  considered  in \cite{ERS}. It was proved  that a sequence $(\rho_N, u_N)$ of point particles (of the form \eqref{pointp})  converges weakly  to a
 weak solution $(\rho,u)$  of the {\it zero pressure gas dynamics} system
\begin{equation}\label{sinai} \frac{\partial\rho}{\partial t} + \frac{ \partial
(u\rho)}{\partial x} = 0 \ \ \ ; \ \ \ \frac{\partial( \rho
u)}{\partial t} + \frac{(\partial \rho u^2)}{\partial x} = 0 \ ,
(x,t)\in (-\infty,\infty)\times (0,\infty) \end{equation}
provided \eqref{weaklim} is satisfied for $t=0$ (and some
additional, technical, conditions).
\par Apart from being a continuous limit  of the model of point
particles, the zero-pressure gas dynamics \eqref{sinai} attracted
a considerable
 interest by its own. It is an example of  hyperbolic system of a
 pair of
 conservation laws which is degenerate, in the sense
 that the two systems of characteristics coincide (see \cite{Sm}). This
 degeneracy leads to a special type of singular solutions, called
 $\delta-$shocks, studied by several authors.
 \par
 The $\delta-$shock solutions present a challenge for the study of
 \eqref{sinai}, and motivate  the study of measure valued solutions of this
 system, their existence and uniqueness (see, e.g. \cite{B},
 \cite{K}
 \cite{TZ}, \cite{Se} and ref. therein). Unlike the non-degenerate   gas dynamics systems,  the
 entropy condition is not enough to guarantee uniqueness of the
 weak solutions for \eqref{sinai}. However, the evolution  of a finite number of sticking  particles
  is evidently determined uniquely by the initial
 conditions $x_i(0), v_i(0)$. As established in \cite{ERS}, the solution of \eqref{sinai}  is
 unique, as
 a weak limit $N\rightarrow \infty$ of  the point particles dynamics \eqref{pointp},  and depends only on the
  weak limit of the initial data $\rho(x,0),
 u(x,0)$
 (and not on the particular sequence).
\par
It is evident that such a result cannot be extended to higher
space dimension.  Indeed, the collision of a pair of point
particles in the space of dimension $d>1$ is a non-generic event.
This leads to an apparent  paradox. The sticking particle
dynamics, which is a very natural model, cannot converge into a
deterministic macroscopic process in the limit of large particle
numbers, unless the space dimension is one.  A  way to circumvent
such a paradox and obtain, perhaps, a macroscopic limit in higher
dimension  is to replace the assumption of point particles by the
assumptions that the particles posses  a finite size, scaled
appropriately with respect to $N$.
\par
In this paper we attempt to consider the macroscopic limit of a
swarm of  particles of finite size. However, we still restrict
ourselves to particles on the line. We show that a macroscopic
limit exists and is unique in this case, even though such a limit
cannot be described by the zero pressure system \eqref{sinai}. We
also describe this macroscopic limit explicitly. The extension of
this model to higher dimension is a challenge we hope to meet
sometime in the future.
\par
In the rest of this section (section~\ref{mainres})  we describe
the setting of the problem  for swarm of N  particles of finite
size and mass $1/N$, and formulate the main result. Unlike the
case of point particles, there
 is no explicit  Eulerian description of the limit
 $N\rightarrow\infty$, as the zero pressure gas dynamics
 \eqref{sinai} for the system of point particles.
To formulate the limit explicitly  we need a Lagrangian
description of this system. Such a description was introduced in
\cite{BG} for point particles, and takes the form of a scalar
conservation law for the mass cumulation function. This
representation  is reviewed in
 part~\ref{lagdes} of  Section~\ref{MainR} below. In part~\ref{expfor}
 we introduce our main result  in an explicit way (Theorem~\ref{mainth}), taking
 advantage of
 the Lagrangian description.
 \par
 The proof of the main result  involves some extensions of
  elementary results and well known definitions  from convex analysis. For the
 convenience of the reader we collected these definitions and
 results in section~\ref{axil}. The proof of the main result is
 given in section~\ref{proofmain}. The proof of the auxiliary
 results of section~\ref{axil} is given in section~\ref{proofexil}.
\subsection{Point particles of finite size}\label{mainres}
 Consider  $N$ identical
particles of fixed size $\nu$ and mass density $\eps^{-1}$ on the
line. The mass of any each particle is $\nu/\eps$.  We shall
assume a total unit mass, so $ N\nu/\eps=1$.  The density profile
of such a particle whose center is at the origin is given by
$$ h_\nu(x)= \left\{ \begin{array}{cc}
  \eps^{-1}  & |x|\leq \nu/2 \\
  0 & |x|>\nu/2
\end{array}\right. $$
The mass distribution  of the system at time $t$ is described by
the density \begin{equation}\label{rhoNdelta}
\rho_\nu(x,t)=\sum_1^N h_\nu(x-x_i(t)) \ , \end{equation} where
$x_{i+1}(t)\geq x_i(t)+\nu$, $1\leq i\leq N-1$ are the positions
of the particles at time $t$. The velocity field is
\begin{equation}\label{uNdelta} u_\nu(x,t) = \eps\sum_1^N v_i(t)h_\nu(x-x_i(t))
\end{equation} where $v_i(t)$ is the velocity of the $ith$
particle. Particles are assumed to move at constant velocity, as
long as they do not collide. If a pair of particles collides then
they stick together to form a compound particle whose mass and
size is the sum of the corresponding masses and sizes of the
particles before collision. The velocity of the compound particle
after collision is determined by the conservation of linear
momentum, and is constant in time between collisions. This law can
be described as
\begin{multline}\label{colnu} v_i(t+)= \frac{\sum_j
1_\nu\left(x_i(t)-x_{j}(t);|i-j|\right) v_{j}(t) }{\sum_j
1_\nu\left(x_i(t)-x_{j}(t);|i-j|\right) }  \ , \\
x_i(t)=x_i(0)+\int_0^t v_i(s)ds \ . \end{multline} where $
1_\nu(x,y;j) = 1$ if $|x-y|= j\nu$, $1_\nu(x,y;j)=0$ otherwise.

\begin{remark}
Note that \eqref{colnu} implies that the order of the particles on
the line is preserved. Moreover, if the collision time between
particles $i,i+1$ is $t_0$, then these particles are glued to each
other, and move under the same velocity, for {\it any} $t>t_0$.
\end{remark}
\par
The object of this paper is to extend the convergence result of
\eqref{pointp} to a system of particles (\ref{rhoNdelta},
\ref{uNdelta}) of finite, shrinking size $\nu_N\searrow 0$. We
shall prove

\vskip.2in\noindent {\bf Main Result:} {\it The sequence
($\rho_N,u_N$) given by (\ref{rhoNdelta}, \ref{uNdelta}) where
$\nu=\eps/N$ converges weakly, under some additional assumptions
(see Theorem~\ref{mainth}, section~\ref{expfor}), to a pair of
functions $(\rho, u)$ provided \eqref{weaklim} is satisfied at
$t=0$.
 Moreover, $(\rho,u)$ depends only on $\rho(\cdot, 0),
 u(\cdot,0)$.}

\section{Main result} \label{MainR}
\subsection{Explicit formulation of the main result}\label{expfor}
Here we formulate the explicit form of the limit claimed at the
end of Section~\ref{mainres}. Before this we need some new
definitions:
\begin{definition}  \ \ \label{firstd}
  A function $f$ is said to be $\eps-$convex on the
interval $[0,1]$ if $m\rightarrow f(m)-\eps m^2/2$ is convex on
this interval. The set of all $\eps-$ convex functions on $[0,1]$
is called $CON_\eps[0,1]$.
\end{definition}
\begin{definition}\label{conhul}
The $\eps-$convex hull of a function $f$ on $[0,1]$ is
$$ f_\eps(m):= \sup_{\phi\in CON_\eps[0,1]}\left\{  \phi(m) \ ; \ \phi\leq f \right\} $$
\end{definition}
Recall the definition of the Legendre Transform:
\begin{definition} Let   $\Phi:\R\rightarrow \R$.
$\Phi^*:\R\rightarrow \R\cup\{\infty\}$ is the Legendre Transform
of $\Phi$ given by
$$ \Phi^*(m):= \sup_{x\in\R} \{ xm-\Phi(x)\} \ .  $$
\end{definition}
\begin{theorem} \label{mainth} Let a sequence
 ($\rho_N,u_N$) given by (\ref{rhoNdelta},
\ref{uNdelta}) where $\nu=\eps/N$. Let $(\orho_N, \ou_N):=
(\rho_N(\cdot,0), u_N(\cdot,0))$, $(\orho, \ou)$  the weak limit
of $(\orho_N, \ou_N)$. Assume \begin{equation}\label{compact}
supp(\orho_N)\subset K \ \ ; \ \ \|\ou_N\|_\infty < C \ \
N=1,2,\ldots  \ ,
\end{equation} holds
 for  some compact $K\subset \R$ and $C>0$, for any
 $N=1,2,\ldots$. Assume, in addition
 $$ \|\orho\|_\infty < \eps^{-1} \ . $$
   Set
$$ \overline{M}(x)=\int_{-\infty}^x
 \orho(s)ds \ ; \
 \overline{\Phi}(x)=\int_{-\infty}^x
 \overline{M}(s)ds
 \ \ ; \ \ \overline{\Psi}=\overline{\Phi}^* \ , $$
 $$ v(m)=\overline{u}\left( \partial_m\overline{\Psi}\right) \  ; \ \ \oV(m)=\int_0^m v(s)ds \ , $$
 \begin{equation}\label{hjsol3}\Psi(m,t)= \left( \overline{\Psi} + t
\overline{V}\right)_\eps(m) \  , \ \Phi(,t):= \Psi^* (,t) \ .
\end{equation} Then \begin{equation}\label{finres}
\rho(x,t)=\partial^2_x\Phi(x,t) \ \ ; \ \ u(x,t)=
v\left(\partial_x\Phi(x,t)\right) \ . \end{equation} is the weak
limit of $(\rho_N, u_N)$.
\end{theorem}
\subsection{Lagrangian description}\label{lagdes}
 In \cite{BG} the weak solution of \eqref{sinai} was interpreted in
terms of Lagrange coordinates.  Let  $M=M(x,t)$ be an entropy
solution of the scalar conservation law.
\begin{equation}\label{conlaw} \frac{\partial M}{\partial t} +
\frac{\partial \oV (M)}{\partial x}=0  \  \ , \ M(x,0)=\oM(x).
\end{equation} Equation \eqref{conlaw} can be integrated once.
Setting \begin{equation}\label{set}\Phi(x,t)=\int_{-\infty}^x
M(s,t)ds ,
\end{equation} \eqref{conlaw} takes the form of the
Hamilton-Jacobi equation \begin{equation}\label{hj} \frac{\partial
\Phi}{\partial t} + \oV\left( \frac{\partial \Phi}{\partial
x}\right)=0 \ \ ; \ \ \Phi(x,0)=\overline{\Phi}(x):=
\int_{-\infty}^x \oM_(s)ds\end{equation}

  The {\it viscosity solution} (see, e.g. \cite{E}) of \eqref{hj}  is given
by \begin{equation}\label{hjsol1}\Phi(x,t)= \left( \overline{\Psi}
+ t \oV\right)^*(x) \  \end{equation}  where
$\overline{\Psi}=\overline{\Phi}^*$. Since $\Phi$  is a convex
function of $x$ for any $t$ by definition, $\partial^2_x\Phi$
exists as a measure on $\R$ for any $t>0$. Taking the Legendre
transform of $\Psi$ we obtain
\begin{equation}\label{hjsol2}\Psi(m,t)= \left( \overline{\Psi} +
t \oV\right)^{**}(m) \  \end{equation} where $\Psi=\Phi^*$ for any
fixed $t$.
\par
It is shown in \cite{BG} that the weak solution of \eqref{sinai}
satisfies
$$ \rho(x,t)dx=\partial^2_x\Phi(x,t) \ \ ; \ \ u(x,t)=
\oV^{'}\left(\partial_x\Phi(x.t)\right) \ . $$
\begin{remark}
Recall that for any function $f:[0,1]\rightarrow \R$, the convex
hull $f_0$ is defined by
$$ f_0(m):= \sup_{\phi\in CON_0[0,1]}\left\{  \phi(m) \ ; \ \phi\leq f \right\} $$
where $CON_0([0,1]$ is the set of all convex functions on $[0,1]$
(consistent with Definition~\ref{firstd}) . The convex hull of a
function is obtained by applying twice the Legendre transform: $
f^{**}= f_0$, so \eqref{hjsol2} can be written as
\begin{equation}\label{hjsol4}\Psi(m,t)= \left( \overline{\Psi} + t
\oV\right)_0(m) \  . \end{equation} Compare \eqref{hjsol4} to
\eqref{hjsol3}.
\end{remark}
\section{Auxiliary results and definitions}\label{axil}
Recall Definitions \ref{firstd},\ref{conhul}.
\begin{lemma}\label{equiv} The following conditions are equivalent
\begin{description}
\item{i)} $f\in CON_\eps[0,1]$.
\item{ii)} for any $m_1<m_2$ in $[0,1]$ and any $s\in [0,1]$,
$$ f\left(s m_1+(1-s)m_2\right) \geq sf(m_1) + (1-s)f(m_2) + \eps
s(s-1)/2 \ . $$
\item {iii)} $f= f_\eps$.
\end{description}
\end{lemma}
\begin{definition}\label{epspar}
An $\eps-$parabola is the graph of  a quadratic function $P=P(s)$
on $[0,1]$  where $d^2P/ds^2\equiv \eps$.
\end{definition}

\begin{remark}
Lemma~\ref{equiv}-(ii) can be stated as follows:  $f\in
CON_\eps[0,1]$ if and only if the graph of $f$ on any interval
$(m_1,m_2)\subset [0,1]$ is below any $\eps-$parabola on the same
interval which connect the points $(m_1, f(m_1))$ and $(m_2,
f(m_2))$.
\end{remark}
From this remark  we can easily obtain the following:
\begin{corollary}\label{!!!}
Let $f\in CON_\eps[0,1]$  and $0\leq m_1<m_2\leq 1$. Let $P$ be an
$\eps-$parabola connecting $ (m_1, f(m_1))$ and $(m_2, f(m_2))$.
Define
$$ g(m):= \left\{ \begin{array}{cc}
  P(m) & \text{if} \ m_1\leq m \leq m_2 \\
  f(m) & \text{otherwise}
\end{array}\right. \ .
$$
Then $g\in CON_\eps[0,1]$ as well.
\end{corollary}
\begin{corollary}\label{depscon}
If $f$ is sequentially $C^2$, convex function which satisfies
$f^{''}\geq \eps$ at all but a countable number of points, then
$f$ is $\eps-$convex.
\end{corollary}
\begin{definition}\label{clusex} Let $\Psi$ be a continuous function on $[0,1]$.
 \begin{itemize}
 \item A point
$m\in[0,1]$ is called an $\Psi-$cluster point if there exists
$m_1<m<m_2$  so that \begin{equation}\label{exdef} s \Psi(m_1) +
(1-s)\Psi(m_2) \leq \Psi\left(sm_1+(1-s)m_2\right) - \eps
s(1-s)/2\end{equation} holds for any $s\in[0,1]$.
 \item
 A point
$m\in[0,1]$ is called an $\Psi-$exposed point if for any $m_1< m <
m_2$ \begin{equation}\label{xe} \Psi(m) <\frac{m_2-m}{m_2-m_1}
\Psi(m_1) + \frac{m-m_1}{m_2-m_1} \Psi(m_2)- \eps
\frac{(m_2-m)(m-m_1)}{(m_2-m_1)^2} \ . \end{equation}
\item The set of all cluster points of $\Psi$ is denoted by $C_\Psi$. The set of all exposed  points of $\Psi$ is denoted by
$E_\Psi$. It's closure in $[0,1]$ is $\overline{E}_\Psi$.
\end{itemize}
\end{definition}
\begin{remark}
\begin{description}
\item {i)} \ Condition\eqref{exdef} states that the graph of $\Psi$  at $(m_1,m_2)$ is below
the $\eps-$parabola connecting $(m_1, \Psi(m_1))$ and $(m_2,
\Psi(m_2))$.
 \item{ii)} Condition
\eqref{xe} can be stated as follows: The point $(m,\Psi(m))$ is
below the $\eps-$parabola connecting  the points $(m_1,
\Psi(m_1))$ and $(m_2, \Psi(m_2))$.
\end{description}
\end{remark}
By Lemma~\ref{equiv}-(ii) we obtain
\begin{corollary}\label{pcon}
If $\Psi\in CON_\eps[0,1]$ then the equality holds in
\eqref{exdef}. In particular,  $\Psi$ coincides with an
$\eps-$parabola on any interval contained in $C_\Psi$.
\end{corollary}

\begin{lemma}\label{equiv1}
\begin{enumerate}
\item\label{open}
The set of cluster points  $C_\Psi$ is open.
\item\label{comp} $[0,1] = C_\Psi\cup \overline{E}_\Psi$ and $C_\Psi\cap \overline{E}_\Psi=\emptyset$.
\item\label{XepsX} If $\Psi_\eps < \Psi$ on some interval $(m_1,m_2)$,
then $\Psi_\eps$ coincides with an $\eps-$parabola on $(m_1,m_2)$.
\item \label{extrem} If $m\in\overline{E}_{\Psi_\eps}$ then
$\Psi_\eps(m)=\Psi(m)$.
\item\label{c=c} $C_\Psi= C_{\Psi_\eps}$. In particular,
$\overline{E}_{\Psi_\eps}=\overline{E}_\Psi$.
\item\label{determined} If $m\in\overline{E}_{\Psi}$ then
$\Psi_\eps(m)=\Psi(m)$. Moreover, the  function $\Psi_\eps$ is
determined everywhere by $\Psi$ on $E_\Psi$.
\item \label{add} If $(m_1,m_2)\subset C_\Psi$ is a maximal
interval\footnote{That is, if $ J\supseteq (m_1,m_2)$ is an
interval contained in $C_\Psi$, then $J=(m_1,m_2)$.} of $C_\Psi$,
 then \eqref{exdef} holds for any $m\in (m_1,m_2)$.
\end{enumerate}
\end{lemma}

\begin{lemma}\label{onlyE}
$m\in E_\Psi$ if and only if \eqref{xe} holds for any $m_1< m<
m_2$ which satisfy $m_i\in \overline{E}_\Psi$, $i=1,2$.
\end{lemma}

\begin{definition}\label{defV}
If $V$ is a continuous function  and $\Psi$ is absolutely
continuous
on $[0,1]$, then $V_\Psi$ is a continuous function defined by: \\
$V_\Psi(m):= V(m)$  if $m\in \overline{E}_\Psi $, $V_\Psi$ is a
linear function on any interval $(m_1, m_2)\subset C_\Psi$.
\end{definition}
We now define the {\it propagator} of a $\eps-$convex function
$\Psi$ on $[0,1]$, {\it given} $V$:
\begin{definition}
If $\Psi$ is $\eps-$convex and $V$ absolutely continuous function
 on $[0,1]$ , then ${\cal F}^V_{(t)}[\Psi]:= \left[ \Psi + t
V_\Psi\right]_\eps$.
\end{definition}
\begin{lemma}\label{cineq}If $t>\tau$ then $C_{{\cal F}^V_{t}[\Psi]} \supseteq C_{{\cal
F}^V_{\tau}[\Psi]}$. In particular, $\overline{E}_{{\cal
F}^V_{t}[\Psi]} \subseteq \overline{E}_{{\cal F}^V_{\tau}[\Psi]}$
by Lemma~\ref{equiv1}-(\ref{comp}).
\end{lemma}

 We now claim  the semigroup property of
${\cal F}^V$:

\begin{proposition}\label{thmain} Given a continuous $V$ and $\eps-$convex function $\Psi$ on $[0,1]$,  for any  $t\geq\tau\geq 0$
\begin{equation}\label{th1}{\cal F}^V_{(t-\tau)}\left[{\cal
F}^V_{(\tau)}[\Psi]\right] = {\cal F}^V_{(t)} [\Psi] \ .
\end{equation}
\end{proposition}
Finally, we shall need the following:
\begin{lemma}\label{conv}
 If $\{\Psi_N\}$ is a sequences of  continuous function which converges uniformly to
$\Psi$ on $[0,1]$, then $\left\{ [\Psi_N]_\eps\right\}$ converges
uniformly to $[\Psi]_\eps$.
\end{lemma}
\section{Proof of the Main Theorem}\label{proofmain} Set $$\orho_N:=\rho_N(,0) \ \
, \ \ \overline{u}_N:=u_N(,0) \ , \ \ M_N(x,t):=\int_{-\infty}^x \
\rho_N(s,t)ds
$$  and $X_N(m,t)$ the generalized inverse of $M_N$ as a function of $x$. Let $$ \tPsi^{(N)}(m,t):=\int_0^m X_N(s,t)ds
\ \ , \ \
 \oPsi^{(N)}(m):=\tPsi^{(N)}(m,0) $$
 \begin{equation}\label{tildePhi} \ntPhi(x,t)= \left( \ntPsi\right)^*(x,t) = \int_{-\infty}^x M_N(s,t)ds \
 , \ \  \overline{\Phi}^{(N)}(x):=\tPhi^{(N)}(x,0) \ .
 \end{equation}
\begin{equation}\label{lawofcol}v_N(m)=
\overline{u}_N\left(\partial_m\oPsi^{(N)}(m)\right) \ ; \ \
\oV^{(N)}(m)=\int_0^m v_N(s)ds \  \end{equation} and
$$ \Psi^{(N)}(m,t):= {\cal
 F}^{(N)}_{(t)}\left[\oPsi^{(N)}\right](m)   \ ; \ \ \Phi^{(N)}(x,t)=\left(\Psi^{(N)}\right)^*(x,t) \ ,$$
 where
 $$ {\cal F}^{(N)}_{(t)}:= {\cal F}^{\oV^{(N)}}_{(t)} \ . $$
 By \eqref{tildePhi}, \eqref{lawofcol} and the law of collision \eqref{colnu}  we obtain
\begin{equation}\label{rhoN} \partial^2_x\ntPhi = \partial_x M_N = \rho_N  \ \
; \ \ v_N\left( \partial_x\ntPhi\right) = u_N  \ . \end{equation}
 Our object is to show
 \begin{equation}\label{1} \tPsi^{(N)}=\Psi^{(N)}  \ \ \text{i.e} \ \ \tPhi^{(N)}=\Phi^{(N)} \end{equation}
 for any $N$, and that
 \begin{equation}\label{2} \lim_{N\rightarrow \infty} \ntPhi(,t) = \Phi(,t) \ \ , \ \ res. \ \
  \lim_{N\rightarrow \infty} \ntPsi(,t) = \Psi(,t)\end{equation}
 exists locally uniformly on $\R$ (res. uniformly on $[0,1]$)  for any $t>0$.
\par
Granted \eqref{1} and \eqref{2}, we can prove the Theorem as
follows: \par\noindent
 By assumption \eqref{compact} and the law
of collisions \eqref{colnu}, the supports of $\rho_N(,t)$ are all
contained in a compact set $K_t\subset \R$. The limit $\Phi$ in
\eqref{2} is clearly a convex function and defines a density
$\rho=\partial^2_x\Phi$ of a probability measure  for any $t\geq
0$, which is also supported in $K_t$.  By \eqref{rhoN} and
\eqref{2} it follows that $\rho$ is the weak limit of $\rho_N$ for
any $t\geq 0$.
\par
Now, the limit \begin{equation}\label{limx1}
\lim_{N\rightarrow\infty} \int_{-\infty}^\infty
\ou_N(x)\orho_N(x)\phi(x) dx=\int_{-\infty}^\infty
\ou(x)\orho(x)\phi(x)\end{equation} holds by assumption. We change
the variable $x$ into $m=\partial_x\overline{\Phi}^{(N)}$. Recall
that $x=\partial_m\overline{\Psi}^{(N)}$ is the inverse relation
and using $\orho_N=\partial^2_x\overline{\Phi}^{(N)}$,
$\orho=\partial^2_x\overline{\Phi}$  and \eqref{lawofcol} we
rewrite \eqref{limx1} as
\begin{equation}\label{limm1}\lim_{N\rightarrow\infty}\int_{-1}^1
v_N(m)\phi\left(\partial_m\overline{\Psi}^{(N)}\right) dm=
\int_{-1}^1 v(m)\phi\left(\partial_m\overline{\Psi}\right) dm \ .
\end{equation} From \eqref{2} evaluated at $t=0$ we obtain $
\overline{\Phi}^{(N)}\rightarrow\overline{\Phi}$ locally uniformly
on $\R$. Since $\overline{\Psi}^{(N)}$ (res. $\overline{\Psi}$)
are the Legendre transforms of $\overline{\Phi}^{(N)}$ (res.
$\overline{\Phi}$), it also follows that $
\overline{\Psi}^{(N)}\rightarrow\overline{\Psi}$ uniformly in
$[0,1]$. Moreover, since $\overline{\Psi}^{(N)}$ are convex it
follows that $\partial_m\overline{\Psi}^{(N)}\rightarrow
\partial_m\overline{\Psi}$ strongly in
$L^1[0,1]$. Recall that the sequence $v_N$ is uniformly bounded in
$L^\infty [0,1]$ by assumption. From this,  \eqref{limm1} and the
obtained $L^1$ convergence
$\partial_m\overline{\Psi}^{(N)}\rightarrow
\partial_m\overline{\Psi}$ we obtain that $v$ is the  {\it
unique} weak $L^\infty$ limit of $v_N$.
\par
We have to prove the existence of $u=u(x,t)$ for which
\begin{equation}\label{limx2}\lim_{N\rightarrow\infty}\int_{-\infty}^\infty
u_N(x,t)\rho_N(x,t)\phi(x) dx=\int_{-\infty}^\infty
u(x,t)\rho(x,t)\phi(x)\end{equation} holds for all $t>0$ and
$\phi\in C_0(\R)$. We note that by  \eqref{tildePhi} and
\eqref{lawofcol} and \begin{equation}\label{undef}
u_N(x,t)=v_N\left(
\partial_x\ntPhi(x,t)\right) \ .
\end{equation} Using the change the variable $x$ into $m=\partial_x\ntPhi$,
recalling that $x=\partial_m\ntPsi$ is the inverse relation and
using \eqref{rhoN},  \eqref{tildePhi} and \eqref{undef}   we write
the left side of \eqref{limx2} as
\begin{equation}\label{limm2}\lim_{N\rightarrow\infty}\int_{-1}^1 v_N(m)
\phi\left( \partial_m \ntPsi(m,t)\right) dm\end{equation} By the
same argument as above we observe, using \eqref{2},  that
$\partial_m\ntPsi(,t)\rightarrow
\partial_m\Psi(,t)$ in  $L^1[0,1]$ for any fixed $t>0$.
Since $\phi$ is continuous it follows that \eqref{limm2} equals
\begin{equation}\label{limm3} \lim_{N\rightarrow\infty}\int_{-1}^1 v_N(m)
\phi\left(
\partial_m \Psi(m,t)\right) dm \ . \end{equation} Since we know already  that
the weak limit in $L^\infty[0,1]$ of $v_N$ is $v$, it follows that
\eqref{limm3} equals
$$ \int_{-1}^1 v(m)
\phi\left(
\partial_m \Psi(m,t)\right) dm  \ ,  $$
which implies that \eqref{limx1} is satisfied were
$$ u(x,t)= v\left( \partial_x\Phi(x,t)\right)  \ . $$
This verifies the second claim in \eqref{finres}.
\par
We now turn to the proofs of \eqref{1} and \eqref{2}: \\
Let $\{ t_l\}$ be the set of  collision times corresponding to
($\rho_N, u_N$). This implies that  there exists  sets $J_l\subset
\{1, \ldots N-1\}$ where \begin{description} \item{i)} \
$x_{i+1}(t)-x_i(t) > \nu_N$ for any $0\leq t<t_l$ and $i\in J_l$.
\item{ii)} \ $x_{i+1}(t_l)-x_i(t_l) = \nu_N$ for any $i\in J_l$.
\item{iii)} \ If $i\not\in J_l$ then either
$x_{i+1}(t)-x_i(t)>\nu_N$ or $x_{i+1}(t)-x_i(t)=\nu_N$ for any
$t_{l-1} \leq t\leq t_l$, were $t_0\equiv 0$.
\end{description}
Now, we observe that  for any  $t\in [t_l,t_{l+1}]$,
\begin{equation}\label{firstind} \ntPsi(,t)= \ntPsi(,t_l) + (t-t_l)
V_{\ntPsi(,t_l)}\end{equation} is $\eps-$convex. To see this, note
that $\partial_m\ntPsi(,t) \equiv X_N(, t)$ is monotone
non-decreasing in $m$ for $t\in[t_l,t_{l+1}]$. Indeed, $X_N(, t)$
is the generalized inverse of $M_N(,t)$ which is monotone by
definition. In addition, any $m\in [0,1]$ which is not an integer
multiple of $\nu_N$ is contained in $C_{\ntPsi(,t_l)}$, so, by
Corollary~\ref{pcon}, $\partial^2_m\ntPsi(m,t)=\eps$ for all but a
finite number of $m$.   It follows that $\ntPsi(,t)$ is
$\eps-$convex by Corollary~\ref{depscon}.
\par
We now proceed to the proof of \eqref{1} by induction. At $t_0=0$
we get
$$ \ntPsi(m,t_0)=\Psi^{(N)}(m,t_0)= \overline{\Psi}^{(N)} (m)$$
by definition. From the $\eps-$convexity of \eqref{firstind} we
obtain
$$ \Psi^{(N)} (m,t) =\left[ \ntPsi(, 0) + t
V_{\ntPsi(,0)} \right]_\eps(m) = \ntPsi(m,t)
$$
for $t_0\leq t\leq t_1$. Suppose now that we verified $\Psi^{(N)}
(m,t)  = \ntPsi(m,t)$ for $t\leq t_j$. Then \eqref{firstind}
implies $$ \ntPsi(,t)= \left[\ntPsi(,t_j) + (t-t_j)
V_{\ntPsi(,t_j)}\right]_\eps := {\cal
F}^{(N)}_{(t-t_j)}\left[\ntPsi(,t_j)\right](m)$$ for $t_j\leq
t\leq t_{j+1}$. But $\ntPsi(,t_j)= \Psi^{(N)}(, t_j) \equiv {\cal
F}^{(N)}_{(t_j)}\left[\overline{\Psi}^{(N)}\right]$ by the
induction hypothesis, so
$$ \ntPsi(m,t)=  {\cal
F}^{(N)}_{(t-t_j)}\left[ {\cal
F}^{(N)}_{(t_j)}\left[\overline{\Psi}^{(N)}\right]\right](m) \ .
$$
This verifies
$$ \ntPsi(,t)= {\cal
F}^{(N)}_{(t)}\left[\overline{\Psi}^{(N)}\right]
\equiv\Psi^{(N)}(,t)$$ for $t_j\leq t\leq t_{j+1}$ by
Proposition~\ref{thmain}.
\par
We now prove \eqref{2}: From the weak convergence
$\orho_N\rightarrow \orho$ we obtain the $L^1[0,1]$ convergence
 $\overline{X}_N:= X_N(,0) \rightarrow \overline{X}$. Since $\overline{X}_N\equiv \partial_m\oPsi^{(N)}$ it
 follows that \begin{equation}\label{psicon}\lim_{N\rightarrow\infty}
 \overline{\Psi}^{(N)}=
 \overline{\Psi}\end{equation}
 uniformly on $[0,1]$ as well.
 \par
  Next, we already proved that $v_N$ converges weakly in
  $L^\infty[0,1]$ to some $v\in L^\infty[0,1]$.
\par\noindent
  This and \eqref{lawofcol} imply that
  \begin{equation}\label{vninfty} \lim_{N\rightarrow\infty} \oV^{(N)}= \oV\end{equation} uniformly on $[0,1]$.

Next, $\partial^2_x\overline{\Phi} = \orho(x) <\eps^{-1}$ by
assumption, so $\partial^2_m\overline{\Psi}   >\eps$ by duality.
This implies that $C_{\overline{\Psi}}=\emptyset$, so
$\oV=\oV_{\overline{\Psi}}$.

Next, we claim  \begin{equation}\label{vninfty1}
\lim_{N\rightarrow\infty} \left\| V^{(N)}_{\overline{\Psi}^{(N)}}-
V^{(N)}\right\|_\infty = 0 \ ,\end{equation} which, together  with
\eqref{vninfty}, implies the uniform convergence
\begin{equation}\label{vninfty2} \lim_{N\rightarrow\infty}
V^{(N)}_{\overline{\Psi}^{(N)}} = V_{\overline{\Psi}} \ .
\end{equation}

 From \eqref{psicon} and \eqref{vninfty2}
   we obtain the uniform convergence of
$$ \lim_{N\rightarrow\infty} \left(\oPsi^{(N)} + t
  \oV^{(N)}_{\oPsi^{(N)}}\right) =\oPsi + t
  \oV_{\oPsi}$$

   By Lemma~\ref{conv} it
  follows that
  $$ \Psi^{(N)}(m,t) \equiv \left[\oPsi^{(N)} + t
  \oV^{(N)}_{\oPsi^{(N)}}\right]_\eps(m)\rightarrow \left[\oPsi + t
  \oV_{\oPsi}\right]_\eps(m)\equiv  \Psi(m,t) \ , $$
  uniformly on $[0,1]$ as well. By \eqref{1} we obtain that
  $$ \lim_{N\rightarrow\infty} \ntPsi(,t)= \Psi(,t)$$
  uniformly on $[0,1]$. This, in turn, implies \eqref{2} by taking
  the Legendre transform of this sequence.

Finally, the claim \eqref{vninfty1} is verified as follows: Let
$m\in [0,1]$. If $m\in E_{\overline{\Psi}^{(N)}}$ then
$V^{(N)}_{\overline{\Psi}^{(N)}}= V_N$.  If  $m\not\in
E_{\overline{\Psi}^{(N)}}$ then  $m\in Supp(\orho_N)$. But, if
$(m_1,m_2)$ is an interval containing $m$ and contained in
$Supp(\orho_N)$, then $(m_1,m_2)$ must contain points {\it not in}
the support of $\orho_N$ for sufficiently large $N$, for,
otherwise, the weak limit $\orho=\eps^{-1}$ on  this interval,
contradiction to the assumption $\|\orho\|_\infty < \eps^{-1}$. In
particular, it follows that for sufficiently large $N$, any such
interval must contain points of $E_{\overline{\Psi}^{(N)}}$, hence
points for which $V^{(N)}=V^{(N)}_{\overline{\Psi}^{(N)}}$. The
sequence $V^{(N)}$ is equi-continuous (since $\partial_mV^{(N)}=
v_N$ and $\|v_N\|_\infty = \|\ou_N\|_\infty < C$ by assumption).
This verifies \eqref{vninfty1}.

\section{Proofs of auxiliary results}\label{proofexil}
The proofs of Lemma~\ref{equiv} and Corollaries \ref{!!!},
\ref{depscon} and \ref{pcon} are rather easy  and we skip it.
\begin{proof} (of Lemma~\ref{equiv1}): \\
Part \eqref{open} is evident from definition.  \\
\eqref{comp} Let $m\not\in C_\Psi$. Let $m_\alpha<m<m_\beta$. By
\eqref{exdef} there exists a point $s_0\in [0,1]$ for which
$$s_0 \Psi(m_\alpha) + (1-s_0)\Psi(m_\beta) +\eps s_0(1-s_0)/2 > \Psi\left(s_0m_\alpha+(1-s_0)m_\beta\right) \ . $$
Since $\Psi$ is continuous, the inequality above holds for some
interval $(s_1,s_2)\subset [0,1]$ where $s_0\in (s_1,s_2)$.  Let
$m_*=s_0 m_\alpha + (1-s_0)m_\beta$ and $m_i=s_i m_\alpha +
(1-s_i)m_\beta$, $i=1,2$. Then $m_*$ satisfies
$$  \Psi(m_*) <\frac{m_2-m}{m_2-m_1} \Psi(m_1) +
\frac{m_*-m_1}{m_2-m_1} \Psi(m_2)- \eps
\frac{(m_2-m_*)(m_*-m_1)}{(m_2-m_1)^2} \ . $$ By \eqref{xe},
$m_*\in E_\Psi$. On the other hand, $m_*\in (m_\alpha, m_2)$ which
is an arbitrary neighborhood of $m$. Hence $m\in\overline{E}_\Psi$
and
\eqref{comp} follows. \\
\eqref{XepsX}  For any $m\in(m_1,m_2)$ let us consider a small
interval $(m_\alpha, m_\beta)\subset (m_1,m_2)$ containing $m$.
Let $P$ the $\eps-$parabola crossing the points $(m_\alpha,
\Psi_\eps(m_\alpha))$ and $(m_\beta, \Psi_\eps(m_\beta))$. Now
consider
$$Y(s):=\left\{ \begin{array}{cc}
  P(s) &  s\in (m_\alpha,m_\beta) \\
  \Psi_\eps(s) & \text{otherwise}
\end{array}\right.$$
Then $Y$ is $\eps-$convex by Corollary~\ref{!!!}. We may choose
the interval $(m_\alpha, m_\beta)$ so small, for which $Y<\Psi$ on
this interval. In particular $Y$ is an $\eps-$convex function
which satisfies $Y\leq \Psi$ on $[0.1]$. Hence $\Psi_\eps\geq Y$
on $[0,1]$ by Definition~\ref{conhul}. On the other hand, since
$\Psi_\eps$ is $\eps-$convex then $\Psi_\eps\leq P$ on $(m_\alpha,
m_\beta)$.It follows that $\Psi_\eps=P$ on $(m_\alpha, m_\beta)$.
The proof follows since the same argument can be applied for any
$m\in (m_1,m_2)$.
\\
 \eqref{extrem}- Let $m\in\overline{E}_{\Psi_\eps}$. Note that
 $\Psi_\eps(m) \leq \Psi(m)$. Suppose $\Psi_\eps(m) < \Psi(m)$.
Let $(m_1,m_2)$ be the maximal interval containing $m$ on which
$\Psi_\eps<\Psi$. In particular, $\Psi(m_i)=\Psi_\eps(m_i)$ for
$i=1,2$. By \eqref{XepsX}, $\Psi_\eps$ coincides with an
$\eps-$parabola on the interval $(m_1,m_2)$. This implies that
$\Psi_\eps$ satisfies condition \eqref{exdef} at $m$,
contradicting $m\in \overline{E}_{\Psi_\eps}$ via point
\ref{comp}.
\\
 \eqref{c=c} -  Let $m\in C_\Psi$, and $m_1,m_2$ as in
\eqref{exdef}. Let $P$ the $\eps-$parabola crossing the points
$(m_1, \Psi(m_1))$ and $(m_2, \Psi(m_2))$. Now consider
$$Y(s):=\left\{ \begin{array}{cc}
  P(s) &  s\in (m_1,m_2) \\
  \Psi_\eps(x) & \text{otherwise}
\end{array}\right.$$
It follows by (***) that $Y$ is an $\eps-$convex function.
Moreover, $Y\leq \Psi$ since  $P\leq \Psi$ on $(m_1,m_2)$ and
$\Psi_\eps\leq \Psi$ everywhere. By Definition~\ref{conhul},
$Y\leq \Psi_\eps$. In particular, $P\leq \Psi_\eps$ on the
interval $(m_1,m_2)$, which implies \eqref{exdef}, so
$C_\Psi\subseteq C_{\Psi_\eps}$.
\\
Conversely, let $m\in C_{\Psi_\eps}$. Set $(m_1,m_2)$ a maximal
interval of $C_{\Psi_\eps}$. Then $m_i\in\overline{E}_{\Psi_\eps}$
for $=1,2$. By point \eqref{extrem} $\Psi_\eps(m_i)=\Psi(m_i)$,
$i=1,2$. Since $\Psi_\eps\leq \Psi$ on $[0,1]$ (in particular, on
$(m_1,m_2)$), and  \eqref{exdef} is satisfied (with an equality)
on $(m_1,m_2)$ for $\Psi_\eps$ by Corollary~\ref{pcon}, it follows
that \eqref{exdef} is also satisfied for $\Psi$ on $(m_1,m_2)$. In
particular $m\in C_\Psi$, so $C_\Psi\subseteq \overline{C}_\Psi$.
\\
\ref{determined} - The first part follows from points \ref{extrem}
and \ref{c=c}. The second part from point \ref{XepsX}.
\\ \ref{add}- Follows from Corollary~\ref{pcon}.
\end{proof}
\begin{proof} (of Lemma~\ref{onlyE}) \\
The "only if" part is trivial from definition. For the "if" part,
let $m_1<m<m_2$, and assume first  that $m_2\in \overline{E}_\Psi$
while $m_1\in C_\Psi$. \par Let $Q$ be the $\eps-$parabola
connecting $(m_1, \Psi(m_1))$ to $(m_2, \Psi(m_2))$. We  show that
$Q(m)> \Psi(m)$. This is equivalent to \eqref{xe} for $m_1,m_2$.
\par
Let $m_\alpha < m_1<m_\beta$ be a maximal interval of $C_\Psi$
containing $m_1$. Since $m\in E_\Psi$ by assumption, then $m_\beta
< m$. Also, $m_\alpha, m_\beta\in\overline{E}_\Psi$ so, by the
assumption of the Lemma, \eqref{xe} is satisfied where $m_1$ is
replaced by $m_\alpha$ or $m_\beta$, respectively.
\par
Now, let $P_\alpha$ be the $\eps-$parabola connecting the points
$(m_\alpha, \Psi(m_\alpha))$ and $(m_2, \Psi(m_2))$. Likewise,
$P_\beta$ is the $\eps-$parabola connecting the points $(m_\beta,
\Psi(m_\beta))$ and $(m_2, \Psi(m_2))$ and $P$  the
$\eps-$parabola connecting the points $(m_1, \Psi(m_1))$ and
$(m_2, \Psi(m_2))$.   Since $m_2\in \overline{E}_\Psi$ and  both
$m_\alpha, m_\beta\in \overline{E}_\Psi$, the condition  of the
Lemma holds for both intervals $(m_\alpha, m_2)$ and $(m_\beta,
m_2)$. It then follows by the assumption of the Lemma that
\begin{equation}\label{x<min}\Psi(m)< \min \left\{ P_\alpha(m),
P_\beta(m)\right\} \ . \end{equation} In addition, $\Psi(m_1)\geq
P(m_1)$ since $(m_\alpha, m_\beta)$ is a maximal interval of
$C_\Psi$ and Lemma~\ref{equiv1}-(\ref{add}) applies.
\\
However, $P(m_1)\geq \min\{ P_\alpha(m_1), P_\beta(m_1)\}$. Hence
$Q(m_1)\geq \min\{ P_\alpha(m_1), P_\beta(m_1)\}$.
 Recalling
that any 2 $\eps-$parabolas may intersect in, at most, one point,
and that $Q(m_2)=P_\alpha(m_2)=P_\beta(m_2)$,  it follows that
$Q(s)\geq \min\{ P_\alpha(s), P_\beta(s)\}$ for $m_2\geq s\geq
m_1$. In particular, $Q(m)> \Psi(m)$ by \eqref{x<min}.
\par
In a similar way we remove the condition $m_2\in
\overline{E}_\Psi$ and prove $Q(m)> \Psi(m)$ for any $m_1<m<m_2$.
This implies $m\in E_\Psi$ by definition.
\end{proof}
\begin{proof} (of Lemma~\ref{cineq}) \\ Let $m\in C_{{\cal F}^V_{\tau}[\Psi]}$. By Lemma~\ref{equiv1}-(\ref{c=c})
$m\in  C_{\Psi+\tau V_\Psi}$. By Definition~\ref{clusex} there
exists an interval $(m_1,m_2)$ containing $m$ where
\begin{multline} s \left[ \Psi(m_1)+\tau V_x(m_1)\right] +
(1-s)\left[\Psi(m_2)+\tau V_\Psi(m_2)\right] \\ \leq\left[
\Psi+\tau V_\Psi\right] \left(sm_1+(1-s)m_2\right)- \eps s(1-s)/2
\ ,
\end{multline}
for any $s\in[0,1]$. That is, \begin{multline}\label{epscon}  \tau
\left[ s
V_\Psi(m_1) + (1-s) V_\Psi(m_2)-V_\Psi\left(sm_1+(1-s)m_2\right)\right]  \\
\leq \Psi
 \left(sm_1+(1-s)m_2\right)-s\Psi(m_1) -(1-s)\Psi(m_2)- \eps s(1-s)/2 \ . \end{multline}
 Since $\Psi$ is $\eps-$convex, the RHS of (\ref{epscon}) is
 non-positive. Hence, the LHS of (\ref{epscon}) is non-positive as
 well. It then follows that if we replace $\tau$ by $t>\tau$ on
 the left of (\ref{epscon}), the inequality will survive. This
 implies that \eqref{exdef} holds for $m_1,m_2$ where $\Psi$ is replaces by $\Psi+tV_\Psi$. Then $m\in C_{\Psi+tV_\Psi}$ as well. The Lemma
 follows since $ C_{\Psi+tV_\Psi}= C_{{\cal F}^V_{(t)}[\Psi]}$ by
 Lemma~\ref{equiv1}-(\ref{c=c}) again.
\end{proof}
\begin{proof} (of Proposition~\ref{thmain}) \\
Set $Y= \Psi + \tau V_\Psi$, $Z=\Psi+tV_\Psi$ and $W=Y_\eps+
(t-\tau)V_{Y_\eps}$.  We shall prove that
\begin{equation}\label{cw=cz} C_W=C_Z \ . \end{equation} Granted
\eqref{cw=cz} we obtain $\overline{E}_Z=\overline{E}_W$ by
Lemma~\ref{equiv1}-(\ref{comp}). Recall that, by
Lemma~\ref{equiv1}-(\ref{extrem}, \ref{c=c}),  if
$m\in\overline{E}_Z$ then $Z(m)=Z_\eps(m)={\cal
F}^V_{(t)}[\Psi](m)$. Also, $m\in \overline{E}_Y$ by
Lemma~\ref{cineq}, so $Y(m)=Y_\eps(m)$, $V_\Psi(m)=V_{Y_\eps}(m)$
hence $W(m)= Y(m)+(t-\tau)V_\Psi(m) = Z(m)$. Hence, by
\eqref{cw=cz} we obtain $Z_\eps(m)=W_\eps(m)$ for any $m\in
\overline{E}_Z=\overline{E}_W$.
 This implies $Z_\eps=W_\eps$
everywhere by Lemma~\ref{equiv}-(\ref{determined}), which implies \eqref{th1}. \\
Proof of  \eqref{cw=cz}:
\\
Let $m\in C_Z$. Let $(m_1,m_2)\subset C_\Psi$  be a maximal
interval of $C_\Psi$.   By Lemma~\ref{equiv1}-(\ref{add})
\begin{multline}\label{Xineq}t\left[  V_\Psi(sm_1+(1-s)m_2)- sV_\Psi(m_1) -(1-s)V_\Psi(m_2)\right]
\\ \leq  [s\Psi(m_1) +(1-s)\Psi(m_2)] -\Psi(sm_1-(1-s)m_2)- \eps s(1-s)/2
\end{multline}
holds for any $s\in [0,1]$.  In turn,
 (\ref{Xineq}) implies
\begin{multline}\label{qqq}(t-\tau)\left[  V_\Psi(sm_1+(1-s)m_2)- sV_\Psi(m_1) -(1-s)V_\Psi(m_2)\right]
\\ \leq   [sY(m_1) +(1-s)Y(m_2)] -Y(sm_1+(1-s)m_2)- \eps s(1-s)/2
\end{multline}
for any $s\in [0,1]$. Since $\overline{E}_Y\subseteq
\overline{E}_\Psi$ by Lemma~\ref{cineq} we obtain that
$V_Y(sm_1+(1-s)m_2)=V_\Psi(sm_1+(1-s)m_2)$ whenever
$sm_1+(1-s)m_2\in \overline{E}_Y$. Moreover,
$Y_\eps(sm_1+(1-s)m_2,\tau)= Y(sm_1+(1-s)m_2)$ under the same
condition.  Since  $(m_1,m_2)$ is assumed a  {\it maximal}
interval of $C_Z$, it follows by Lemma~\ref{equiv1}-(\ref{comp})
that  $m_1,m_2\in \overline{E}_Z$.
 However,
$\overline{E}_Z\subseteq\overline{E}_Y$ (Lemma~\ref{cineq} again),
so $m_1,m_2\in \overline{E}_Y$ and $V_Y(m_i)=V_\Psi(m_i)$,
$Y_\eps(m_i)=Y(m_i)$ for $i=1,2$ as well. It then follows from
(\ref{qqq}) that
\begin{multline}\label{ttt}(t-\tau)\left[  V_Y(sm_1+(1-s)m_2,\tau)- sV_Y(m_1,\tau) -(1-s)V_Y(m_2,\tau)\right]
\\ -   [sY_\eps(m_1,\tau) +(1-s)Y_\eps(m_2,\tau)] +Y_\eps(sm_1-(1-s)m_2,\tau)+ \eps
s(1-s)/2 \geq 0
\end{multline}
holds for any such $s$.  But, on the complement of
$\overline{E}_Y$ in $(m_1,m_2)$, the RHS of (\ref{ttt}) is linear
in $s$. Hence, the inequality (\ref{ttt}) holds for any $s\in
[0,1]$. This, in turn, implies that (\ref{exdef}) is satisfied
where $\Psi$ replaced by $W$. Thus,  $(m_1,m_2)\subset C_W$ so
$C_Z\subseteq C_W$. In particular $\overline{E}_W\subseteq
\overline{E}_Z$.
\par
If $m\in \overline{E}_Z$, then $Z(m)=\Psi(m)+tV_\Psi(m)= Y(m) +
(t-\tau) V_\Psi(m)$. On the other hand, Lemma~\ref{cineq} we know
that $\overline{E}_Z\subset \overline{E}_Y$, so $m\in\overline{
E}_Y$ and by Definition~\ref{defV} we obtain $V_\Psi(m)=V_Y(m)$.
Lemma~\ref{equiv1}-(\ref{determined}) also yields
$Y_\eps(m)=Y(m)$. Hence \begin{equation}\label{z=w}Z(m)= Y_\eps(m)
+ (t-\tau) V_Y(m)=W(m) \ . \end{equation}

Suppose now $m\in E_Z$. Hence, by \eqref{z=w} and
Definition~\ref{clusex},
\begin{equation}\label{con1} W(m)=Z(m)< \frac{m_2-m}{m_2-m_1}
Z(m_1) + \frac{m-m_1}{m_2-m_1} Z(m_2)- \eps
\frac{(m_2-m)(m-m_1)}{2(m_2-m_1)^2} \ .
\end{equation} for {\it any} $m_1<m<m_2$. Suppose, in addition,
$m_i\in\overline{E}_W$, $i=1,2$. Since we know
$\overline{E}_W\subseteq \overline{E}_Z$ then $m_1,m_2\in
\overline{E}_{Z}$.  In particular, $W(m_i)=Z(m_i)$ by \eqref{z=w},
so \eqref{con1} implies
\begin{equation}\label{con2} W(m)< \frac{m_2-m}{m_2-m_1}
W(m_1) + \frac{m-m_1}{m_2-m_1} W(m_2)- \eps
\frac{(m_2-m)(m-m_1)}{2(m_2-m_1)^2} \ .
\end{equation}
Since \eqref{con2} holds for any $m_i\in \overline{E}_W$, it
implies that $z\in E_W$ by Lemma~\ref{onlyE}. This implies
$E_Z\subseteq E_W$ and complete the proof of \eqref{cw=cz}.
\end{proof}

\end{document}